\documentclass[]{pasj00}
\begin{document}
%\SetRunningHead{Author(s) in page-head}{Running Head}
\SetRunningHead{Uchida et al.}{Suzaku X-ray Observation of W44}
\Received{2012/04/04}%{yyyy/mm/dd}
\Accepted{2012/08/01}%{yyyy/mm/dd}

\title{Recombining Plasma and Hard X-ray Filament in \\ the Mixed-Morphology Supernova Remnant W44}

\author{Hiroyuki \textsc{Uchida}\altaffilmark{1}, Katsuji \textsc{Koyama}\altaffilmark{1,2}, Hiroya \textsc{Yamaguchi}\altaffilmark{3}, Makoto \textsc{Sawada}\altaffilmark{1,4}, Takao \textsc{Ohnishi}\altaffilmark{1}, Takeshi Go \textsc{Tsuru}\altaffilmark{1}, Takaaki \textsc{Tanaka}\altaffilmark{1}, Satoshi Yoshiike\altaffilmark{4} and Yasuo Fukui\altaffilmark{4}} %
\altaffiltext{1}{Department of Physics, Graduate School of Science, Kyoto University, Kitashirakawa Oiwake-cho, Sakyo-ku, Kyoto 606-8502, Japan}
\altaffiltext{2}{Department of Earth and Space Science, Graduate School of Science, Osaka University, 1-1 Machikaneyama, Toyonaka, Osaka 560-0043, Japan}
\altaffiltext{3}{Harvard-Smithsonian Center for Astrophysics, 60 Garden St., Cambridge, MA 02138, USA}
\altaffiltext{4}{Department of Physics and Mathematics, Aoyama Gakuin University, 5-10-1 Fuchinobe, Sagamihara, Kanagawa 252-5258, Japan}
\altaffiltext{5}{Department of Physics, Nagoya University, Chikusa, Nagoya 464-8602, Japan}
%\altaffiltext{3}{RIKEN, 2-1 Hirosawa, Wako, Saitama 351-0198, Japan}
\email{uchida@cr.scphys.kyoto-u.ac.jp}


\maketitle

\begin{abstract}
We report new features of the typical mixed-morphology (MM) supernova remnant (SNR) W44. 
In the X-ray spectra obtained with Suzaku, radiative recombination continua (RRCs) of highly ionized atoms are detected for the first time.
The spectra are well reproduced by a thermal plasma in a recombining phase. 
The best-fit parameters suggest that the electron temperature of the shock-heated matters cooled down rapidly from $\sim1$\,keV to $\sim 0.5$\,keV, possibly due to adiabatic expansion (rarefaction) occurred $\sim20,000$ years ago.
We also discover hard X-ray emission which shows an arc-like structure spatially-correlated with a radio continuum filament. 
The surface brightness distribution shows a clear anti-correlation with $^{12}$CO ($J=2$--1) emission from a molecular cloud observed with NANTEN2.
%The origin of the hard X-ray is likely to be synchrotron emission from relativistic electrons accelerated in the vicinity of the cloud. 
%This is the first evidence for presence of TeV electrons in an evolved MM SNR.
While the hard X-ray is most likely due to a synchrotron enhancement in the vicinity of the cloud, no current model can quantitatively predict the observed flux. 
\end{abstract}

\section{Introduction}\label{sec:intro}
Mixed-morphology (MM) supernova remnants (SNRs) are classified based on the characteristic morphology of centrally-filled thermal X-ray emission with a synchrotron radio shell (\cite{Rho98}; \cite{Lazendic06}; \cite{Vink12}).
More than 25\% of the X-ray-detected Galactic SNRs are categorized as the MM type \citep{Jones98}.
Most of the MM SNRs indicate evidence for interaction with dense interstellar medium (ISM), CO and H\emissiontype{I} emissions or OH (1720\,MHz) masers in some cases \citep{Rho98}.
They are also associated with TeV/GeV $\gamma$-ray sources, suggesting the presence of molecular clouds in their vicinity (e.g., \cite{Albert07}; \cite{Aharonian08}; \cite{Hewitt09}; \cite{Abdo10}).
Such ambient environment would play some roles in the evolutions of the morphology and plasma condition in these SNRs; e.g., the central X-ray emission might be enhanced by cool clouds left relatively intact after the passage of blast waves to slowly evaporate in the hot SNR interior \citep{White91}, or density gradients in the pre-existing ISM could contribute the emission profile \citep{Petruk01}.

Recent Suzaku observations revealed the presence of enhanced radiative recombination continua (RRCs) in X-ray spectra of several MM SNRs, IC\,443 \citep{Yamaguchi09}, W49B \citep{Ozawa09}, G\,359.1$-$0.5 \citep{Ohnishi11} and W28 \citep{Sawada12}.
The strong RRC emission can be observable when an ionization temperature ($kT_{\rm z}$\footnote{$kT_{\rm z}$ represents populations of ionization states.
In collisional ionization equilibrium, $kT_{\rm z}$ is equal to the plasma temperature; see also \citet{Masai94}.}) is significantly higher than electron temperature ($kT_{\rm e}$) so that the free-bound transition (recombination process) becomes dominant---hereafter, we call it a recombining plasma (RP).  
These discoveries are dramatically changing our understanding of the SNR evolution, since most young or middle-aged SNRs have either an ionizing plasma (IP: $kT_{\rm e}>kT_{\rm z}$) or a collisional ionization equilibrium plasma (CIE: $kT_{\rm z}=kT_{\rm e}$). 
%Interestingly, all the SNRs where the RPs are discovered are in the MM class. 
While hints of RPs were reported from a few non-MM SNRs (\cite{Hughes03}; \cite{Broersen11}), all the  robust results have been found only in MM SNRs.
It is suggested, therefore, that the formation process of the RPs is closely related to the characteristic environment of the MM SNRs.

The environment may also affect the particle acceleration in SNR blast waves. 
The shock waves propagating into the inhomogeneous ambient medium may produce strong turbulence in the magnetic field, which enhances the efficiency of the particle acceleration in this region (e.g., \cite{Giacalone07}; \cite{Inoue12}).

\begin{table*}[!t]
\caption{Observation logs.}\label{tab:obs}
\begin{center}
\begin{tabular}{lcccccc}
\hline
\hline
& Name & Obs. ID  & Obs. Date & (R.A., Dec.) $_{J2000}$ & ($l$, $b$) &Exposure\\
\hline	   						    		      			  						
Source & W44  & 505004010   & 2010-Apr-10 &  (\timeform{18h56m08s}, \timeform{+01D23'19"})  & (\timeform{34.7D}, \timeform{-0.41D}) & 61.1\,ks\\ 
Background & GR & 500009020   & 2007-Aug-18 &  (\timeform{18h44m01s}, \timeform{+04D04'41"})  & (\timeform{28.5D}, \timeform{+0.21D}) & 98.9\,ks\\ 
\hline
\end{tabular}
\end{center}
\end{table*}

In this paper, we focus on one of the most typical MM SNRs, W44 (also known as G\,34.7$-$0.4 or 3C\,392) to study its X-ray properties and their relation with the environment. 
W44 is a middle-aged ($\sim20,000$\,yr; \cite{Wolszczan91}) SNR located on the Galactic plane at a distance of $D\sim3$\,kpc (Claussen et al.\ 1997).
In the radio band, W44 shows a distorted shell of the synchrotron emission (e.g., \cite{Kundu72}; \cite{Handa87}).
Recent high-resolution radio observations revealed more complicated filaments in the radio continuum shell (\cite{Giacani97}; \cite{Castelletti07}).
Molecular radio emission was detected from the vicinity of W44 by several observations of the CO (\cite{Wootten77}; \cite{DeNoyer79}; \cite{Seta98}; \cite{Reach05}) and H\emissiontype{I} \citep{Koo95} lines. 
More recently, \citet{Seta04} showed a whole view of the high-resolution $^{12}$CO emission map with the 17$\arcsec$-beam of the Nobeyama 45-m radio telescope and reported many hints for the shock-cloud interaction, including cloud evaporation.
The presence of 1720\,MHz OH maser emission is also a sign of the interaction between the SNR shocks and the ambient molecular clouds (e.g., \cite{Claussen97}; \cite{Frail98}).

In the X-ray band, W44 was found to have center-filled morphology by ROSAT observations \citep{Rho94}.
Based on the Chandra result, \citet{Shelton04} proposed that the centrally bright X-ray morphology is caused by either entropy mixing in the form of thermal conductions, or evaporation of swept-up clouds.  
The analyses of combined X-ray spectra of ROSAT, EXOSAT (ME), Einstein (SSS), and Ginga revealed that the plasma was IP and has not reached CIE, and the spectra are interpreted  by cloud evaporation 
(\cite{Rho94}; \cite{Harrus97}).
ASCA obtained higher-quality spectra and determined the ionization temperature ($kT_{\rm z}$) of Si from the flux ratio of K$\alpha$ lines between H-like and He-like ions (H-like/He-like). 
The result was that $kT_{\rm z}$ is nearly equal to the electron temperature $kT_{\rm e}$ \citep{Kawasaki05}, or the thermal plasma was nearly in CIE.

This paper reports revised X-ray features of W44 made with the Suzaku observation.
The bright X-ray flux and apparent X-ray size of $\sim\timeform{0.6D}$ in diameter are best suited to study the detailed plasma structure. 
The high sensitivity and the good energy resolution of Suzaku enable us to determine the spatial structure of the X-rays accurately, in particular, that of the thermal plasma. 
We report the first detection of an RP in this SNR and discovery of hard X-ray emissions. 
The evolution of the RP is quantitatively examined, and then some scenarios for its origin are addressed. 
The hard X-ray structure is compared to that of the molecular cloud with the NANTEN2 observations. 
The origin is discussed in the context of a shock-cloud interaction. 
Throughout this paper, the distance to the SNR is assumed to be 3\,kpc, and the errors quoted are at the 90\% confidence level.

\begin{figure}[!t]
  \begin{center}
   \FigureFile(85mm,85mm){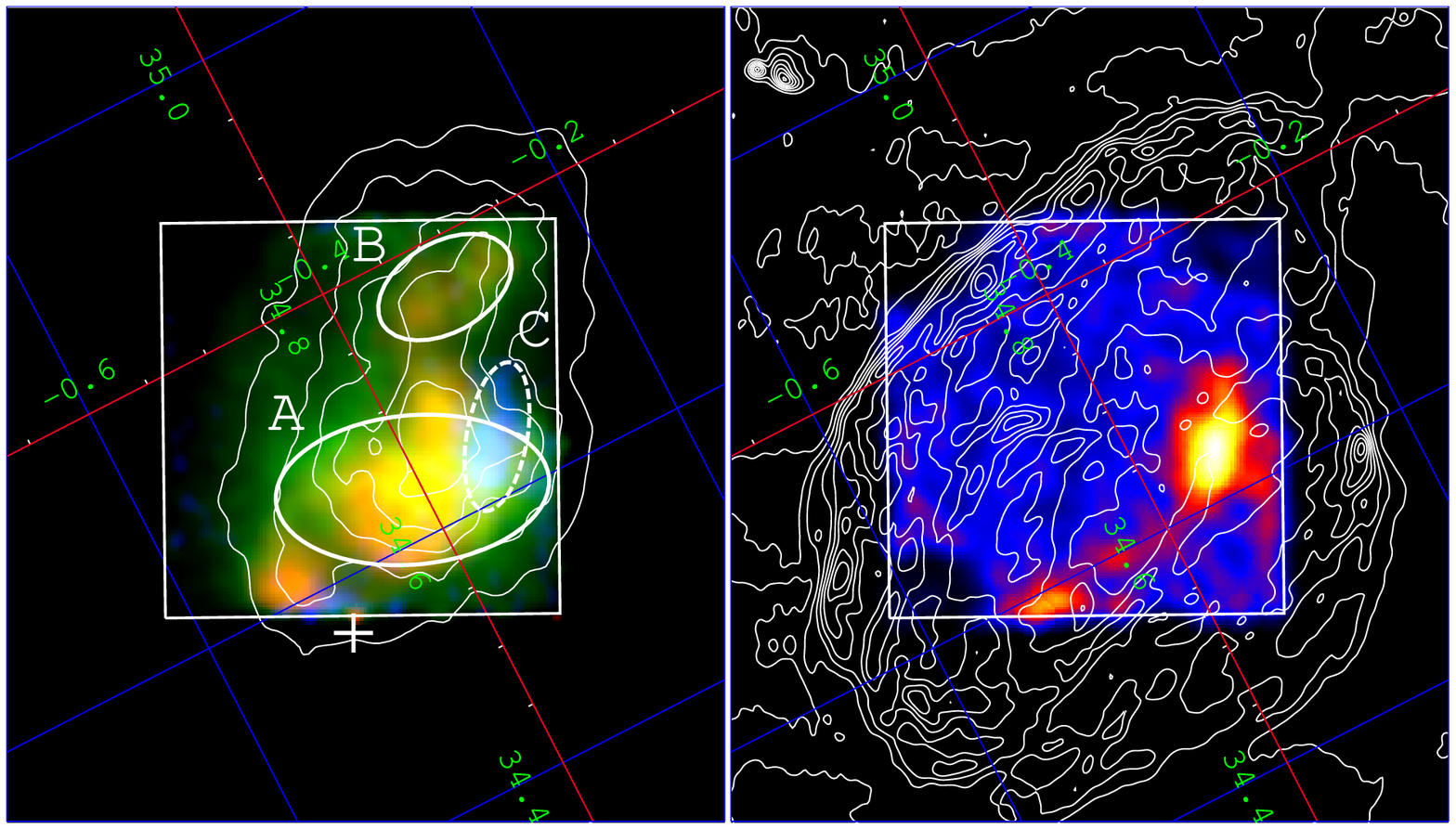}
   \FigureFile(70mm,70mm){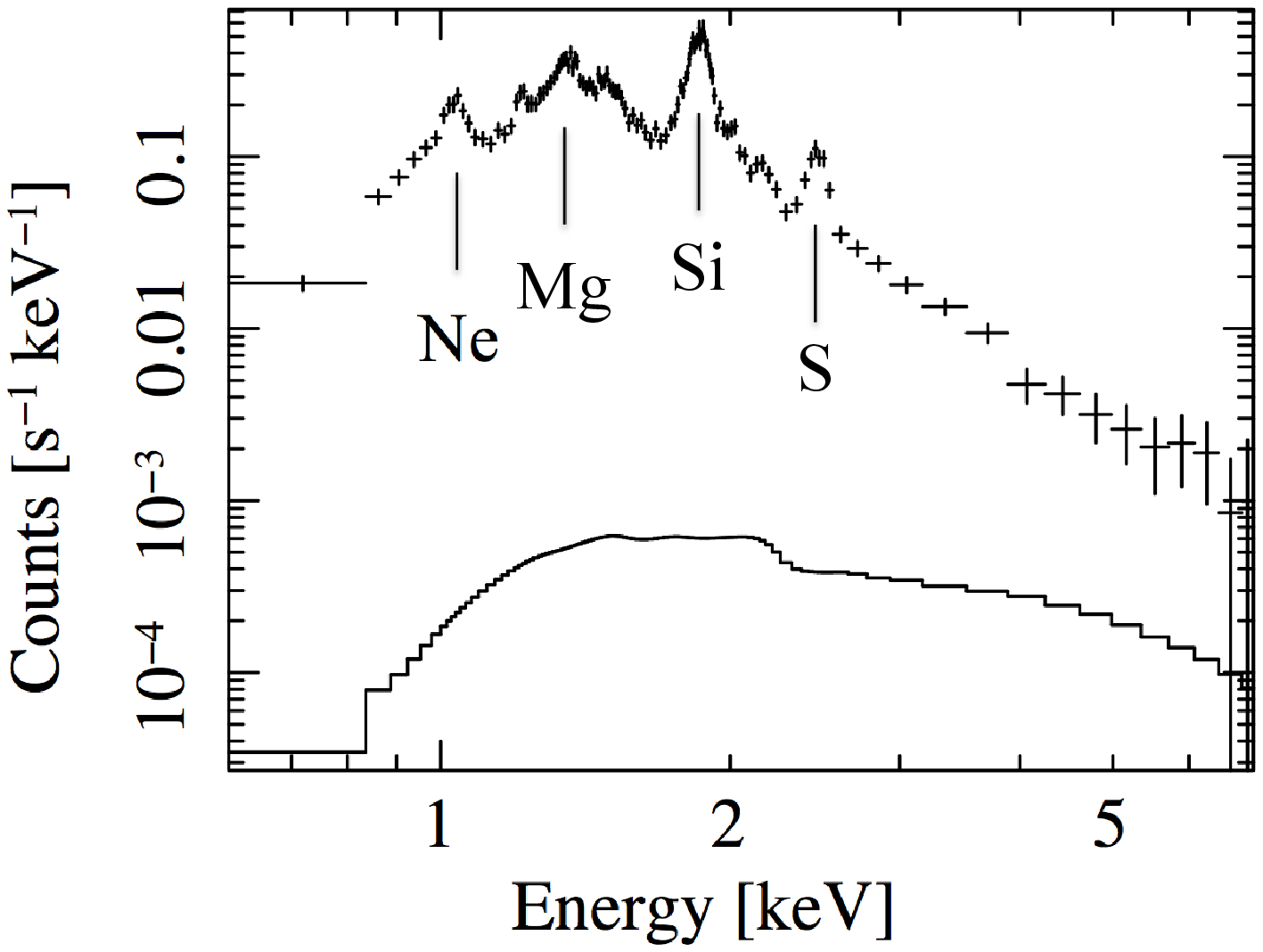}
  \end{center}
  \caption{\textit{Top Left}: Three-color XIS image of W44 overlaid on the Galactic coordinate. The red, green and, blue correspond to the 1.0--2.0\,keV, 2.0--5.0\,keV and 5.0--7.0\,keV bands, respectively. The white contours represent X-ray surface brightness obtained by the ASCA GIS observation \citep{Kawasaki05}. 
 The FoV of the XIS is shown with the white square. The white cross indicates the location of the pulsar PSR\,B1853$+$01. \textit{Top Right}: Hard X-ray band  (5.0--7.0\,keV) image obtained with the Suzaku XIS. The white contours represent the 1442.5\,MHz radio continuum emission with VLA. \textit{Bottom}: Example of the source spectrum extracted from the region C. The solid line is the CXB model spectrum.}\label{fig:image}
\end{figure}

\section{Observations and Data Reduction}
We performed a one-pointing observation of W44 with X-ray Imaging Spectrometer (XIS; \cite{Koyama07}) on board Suzaku. 
The field of view (FoV) covers the X-ray brightest region in this SNR. 
In the southwest part of the FoV, the shock-cloud interaction has so far been confirmed by the CO observations (e.g., \cite{Seta04}). 
The observation log is presented in table \ref{tab:obs}.

The Suzaku observation was performed in April 2010, during the AO5 observing cycle. 
We used data of XIS1 (back-illuminated CCD; BI CCD), and XIS0 and XIS3 (front-illuminated CCDs; FI CCDs). 
XIS2 was unavailable since 2006 November 9, possibly due to a micrometeorite damage.
We used the revision 2.4 of the cleaned event data and combined the $3\times3$ and $5\times5$ pixel events. 
The calibration database (CALDB) updated in September 2011 was used for the data reprocessing. 
We performed data reduction with the version 6.11 of the HEAsoft tools (version 18 of the Suzaku Software).

We also performed observations of the $^{12}$CO ($J=2$--1) emission from 2010 August to 2011 January with NANTEN2, the 4\,m-submillimeter telescope of Nagoya University at Atacama (4865\,m above the sea level) in Chile. 
The half-power beam width at the frequency of the $^{12}$CO ($J=2$--1), 230.580\,GHz, was $\sim\timeform{90"}$.
The observations were made by on-the-fly (OTF) mapping mode and each spatial region was mapped several times in different scanning directions to reduce scanning effects. 
The final pixel size of the gridded data was \timeform{30"}.  
The spectrometer was a digital Fourier transform spectrometer (DFS) with 16384 channels, providing a velocity resolution of 0.08\,km\,s$^{-1}$. 
The root mean square noise per channel after smoothing was $\sim0.3$\,K.

\section{Analysis and Results}
In the following spectral analysis, we used the SPEX software version 2.02.02 \citep{Kaastra96}. 
We employed \textit{xisrmfgen} and \textit{xissimarfgen} to generate the redistribution matrix files (RMFs) and ancillary response files (ARFs), respectively \citep{Ishisaki07}.
The energy band around the neutral Si K-shell edge (1.7--1.8\,keV) is ignored, because an accurate response function in this band was unavailable at the time of the data analysis
\footnote{http://heasarc.nasa.gov/docs/suzaku/analysis/sical.html}.

\subsection{Background Subtraction}
The X-ray images after subtraction of the non X-ray background (NXB) (\textit{xisnxbgen}; \cite{Tawa08}) are given in figure \ref{fig:image}.
The left and right panels show the three-color (three-energy bands) and hard X-ray images of W44, respectively.
To make the source spectra, we further need to subtract the cosmic X-ray background (CXB) and the Galactic ridge X-ray emission (GRXE). 

The major component of the X-ray background is GRXE. 
Since the X-ray emission of W44 extends more widely than the FoV (figure \ref{fig:image}), background data were not taken from the same FoV.  
We used the nearby sky on the Galactic ridge (hereafter GR). 
The observation log for the GR is also shown in table \ref{tab:obs}.
The GR observation was aimed \timeform{6.2D} and \timeform{0.2D} apart from W44 along the Galactic longitude and latitude, respectively. 
This separation cannot be ignored for the background estimation, since the GRXE flux depends largely on the position of the Galactic plane, espetially on the Galactic latitude.  
The spatial distribution of the GRXE flux is modeled by a two-exponential function with scale lengths of 2500\,pc (\timeform{17D}) and 45\,pc (\timeform{0.3D}) along the Galactic longitude and latitude, respectively (\cite{Uchiyama11}).
Therefore, the GRXE flux around W44 is estimated to be $\sim40$\% of that at the GR position.
We subtracted normalized GRXE spectrum, resulting no significant excess above 6.0\,keV in the source spectra of the regions A and B. 
This is consistent with the imaging analysis (figure~\ref{fig:image}) where the no excess was found in the hard X-ray band, with an exception of the region C. 
We hence conclude that the GRXE is properly subtracted.  
The hard X-ray excess in the region C is separately discussed later.

The surface brightness of the CXB is less than $\sim5$\% of that of the GRXE. Therefore, possible spatial fluctuation in its flux can be negligible compared to the GRXE.
We applied a power-low CXB model with a photon index of 1.4, following the result by \citep{Kushino02}.

\subsection{Fit with Thermal Plasma Models}\label{sec:ion}
\begin{figure}[t]
  \begin{center}
   \FigureFile(85mm,85mm){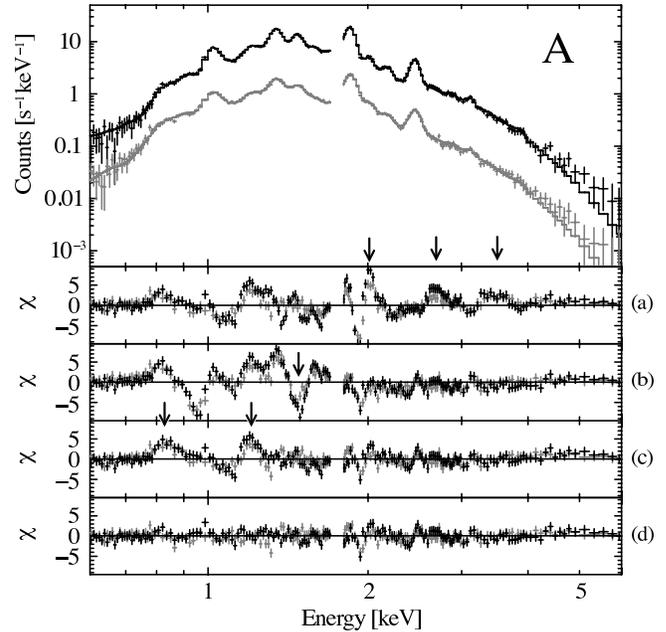}
  \end{center}
  \caption{The top panel shows the region A spectrum fitted with the NEIJ plus two-Gaussian model (see text). Black and gray represent the FI and BI data, respectively. The count rate of FI is multiplied by a factor of 10. The lower panels represent the residuals from the models of CIE (a), RP (b), NEIJ (c), and NEIJ plus 
  two Gaussians (d).}\label{fig:spec}
\end{figure}

In figure \ref{fig:image}-left, we see a bright centrally-filled structure in the soft X-ray band (1.0--5.0\,keV). 
For spectral studies, we selected three representative regions, A, B, and C, in figure \ref{fig:image} left; the former two regions exhibit no hard X-ray emission, while the remaining one shows clear excess in 
the hard X-ray band. 
From the region A, the overlapped ellipse of the region C was excluded.

\begin{figure*}[t]
  \begin{center}
   \FigureFile(55mm,55mm){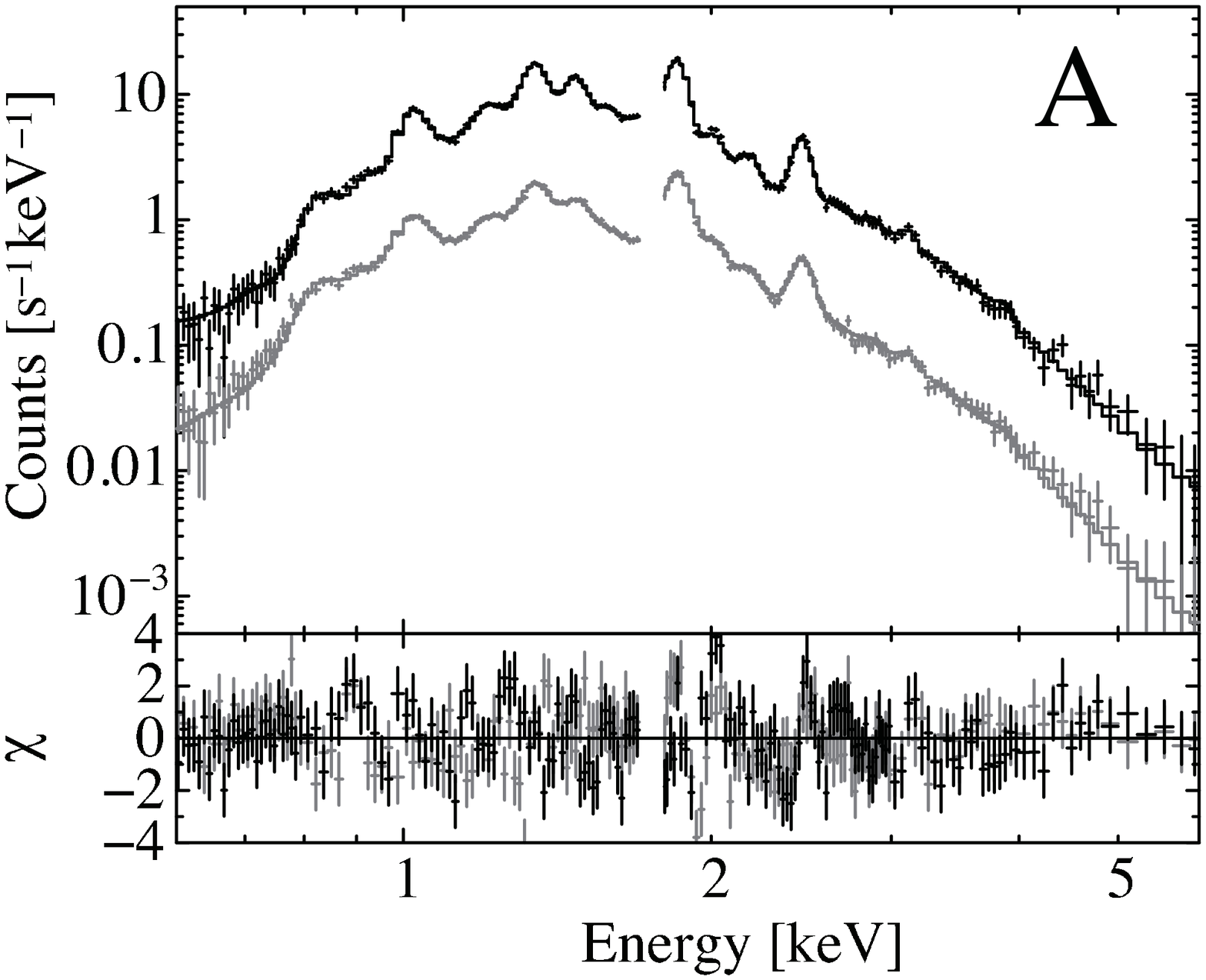}
   \FigureFile(55mm,55mm){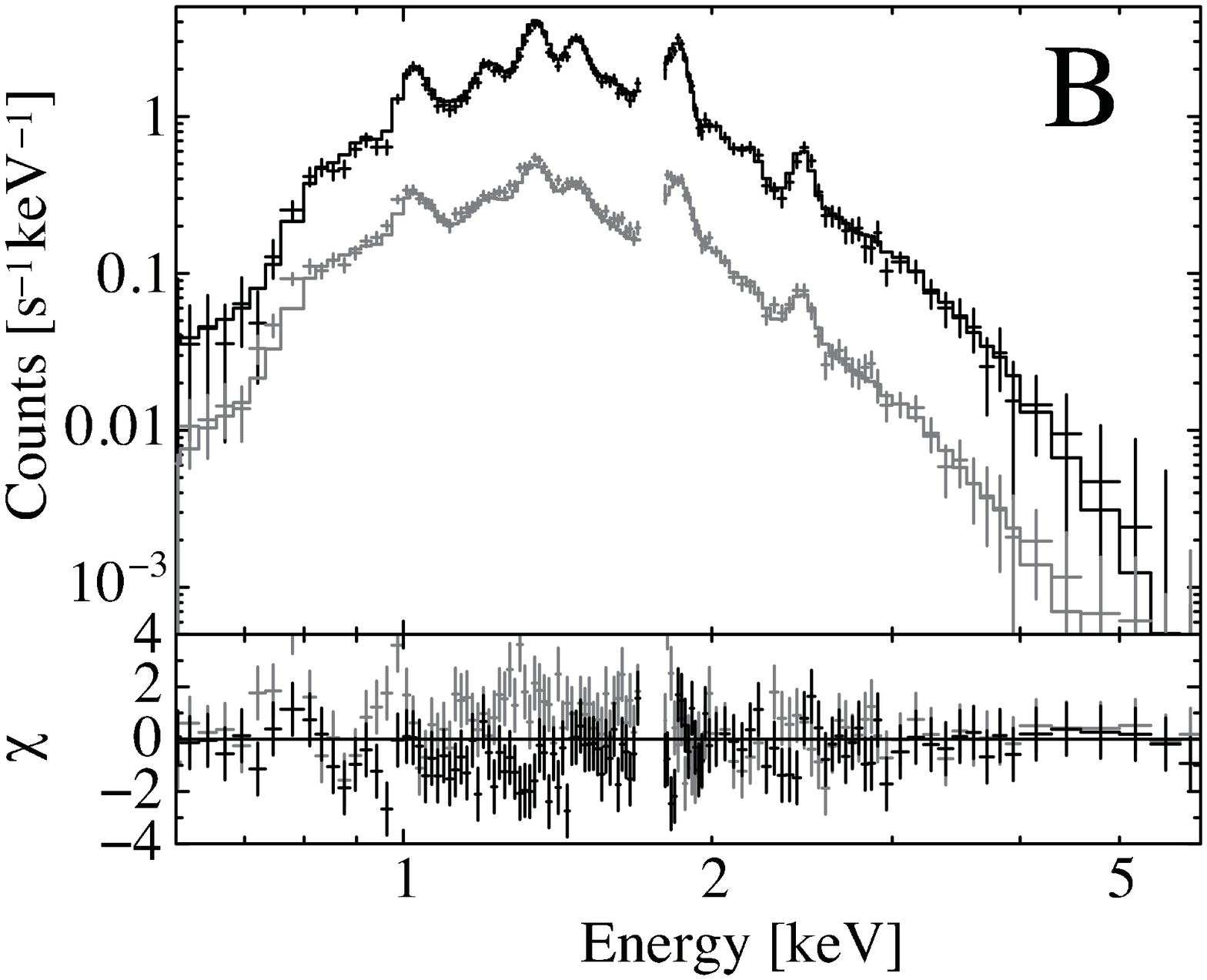}
   \FigureFile(55mm,55mm){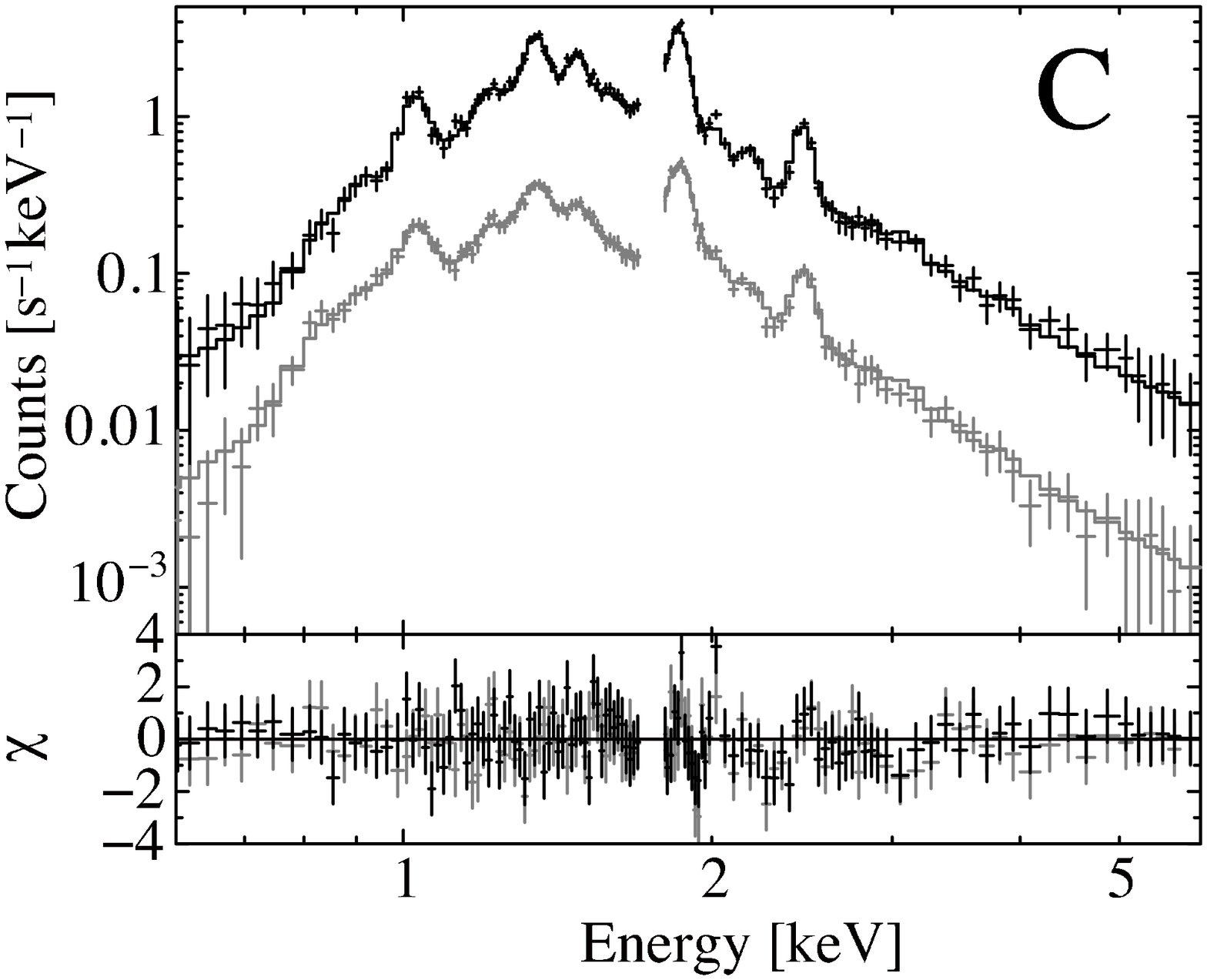}
  \end{center}
  \caption{Spectra of the regions A, B and C fitted with the NEIJ ($+$ two Gaussians)  plus power-law model (see text). Black and gray correspond to the FI ($\times$10) and BI data, respectively. The residual is shown in each lower panel.}\label{fig:spec2}
\end{figure*}

\begin{table*}[t]
  \caption{Best-fit parameters (see figure \ref{fig:spec2})}\label{tab:spec}
  \begin{center}
    \begin{tabular}{llccc}
       \hline 
      \hline
Components & Parameters & \multicolumn{3}{c}{Value} \\
      \hline
 & & region A & region B & region C \\
            Thermal Component (NEIJ) \dotfill		& $kT_{\rm e1}$ [keV]	& 1.00$\pm0.05$		& 0.92$\pm0.03$		& 0.95$^{+0.05}_{-0.06}$ 	\\
								& $kT_{\rm e2}$ [keV] 			& 0.47$\pm0.01$		& 0.49$^{+0.05}_{-0.06}$		& 0.50$\pm$0.02 			\\
%			& $n_{\rm e}t$ [10$^{11}$ s cm$^{-3}$] 					& 6.43$^{+2.96}_{-0.40}$ 	& 5.83$^{+2.42}_{-2.84}$ 	& 7.04$^{+1.66}_{-3.07}$ 	\\
			& log($n_{\rm e}t \ \rm{[s\,cm^{-3}]}$) 					& 11.79$^{+0.06}_{-0.03}$ 	& 11.77$^{+0.15}_{-0.29}$ 	& 11.85$^{+0.09}_{-0.25}$ 	\\
			& $EM^*$ [10$^{58}$ cm$^{-3}$] 			& 42.6$^{+1.3}_{-2.2}$ 		& 28.7$^{+6.1}_{-5.1}$ 		& 26.2$^{+2.2}_{-3.2}$ 		\\
\ \ Abundance (solar)$^\dagger$	   	& Ne		 		& 1.02$^{+0.04}_{-0.05}$ 	& 1.29$^{+0.48}_{-0.32}$ 	& 1.27$^{+0.16}_{-0.40}$\\
    								& Mg	       		& 1.20$^{+0.05}_{-0.06}$ 	& 1.26$^{+0.27}_{-0.16}$ 	& 1.46$^{+0.21}_{-0.13}$ \\
 								& Si		        		& 1.68$^{+0.05}_{-0.07}$ 	& 0.98$^{+0.28}_{-0.14}$ 	& 2.15$^{+0.40}_{-0.22}$ \\
      								& S	          		& 2.02$^{+0.10}_{-0.16}$ 	& 1.01$^{+0.48}_{-0.27}$ 	& 2.59$^{+0.37}_{-0.60}$ \\
       								& Ar (=Ca)		& 1.91$^{+0.34}_{-0.30}$		& $<$1.23 				& $<$1.09 			\\
     								& Fe (=Ni)			& 0.05$\pm0.01$ 		 	& 0.06$^{+0.05}_{-0.06}$  	& $<$0.07 		 	\\
					\\
Powerlaw Component\dotfill    					& $\Gamma$ 		& 2.12 (fixed) 				& 2.12 (fixed) 				& 2.12$\pm$0.11 \\
%& $\textit{Norm}$ [10$^{43}$ photons s$^{-1}$\,keV$^{-1}$] & 0.36$^{+0.22}_{-0.21}$		& $<$0.32 				& 3.20$^{+0.52}_{-0.97}$ \\
& flux [$10^{-12}$\,ergs\,cm$^{-2}$\,s$^{-1}$] 				& $0.15\pm0.09$			& $<0.14$					& $1.37^{+0.22}_{-0.42}$ \\
\\
Absorption\dotfill & \textit{N}$\rm _H$ [10$^{22}$ cm$^{-2}$] 	& 1.3$\pm$0.2 				& 1.2$\pm$0.2				& 1.3$\pm$0.2 \\
\hline
&  $\chi ^2$/dof  									& 525/347  				& 323/219 				& 202/237  \\
      \hline
\multicolumn{2}{l}{$^*$ Volume emission measure at the distance of 3\,kpc.}\\
\multicolumn{2}{l}{$^\dagger$ The other elements not listed here are fixed to solar values.}\\
\multicolumn{2}{l}{The errors are in the range $\Delta\chi^{2}<2.7$ on one parameter.}\\
    \end{tabular}
 \end{center}
\end{table*}

We first analyze the spectrum of the region A. 
The previous observations suggested that the X-ray spectrum of the SNR's center is characterized by a thermal plasma in either ionizing state or CIE (e.g., \cite{Harrus97};  \cite{Kawasaki05}). 
Therefore, we applied a single-component CIE model with free parameters of plasma temperature $kT_{\rm e}$($=kT_{\rm z}$),  emission measure ($EM=\int n_{\rm e}n_{\rm H} V$), and column density $N{\rm_H}$ for interstellar absorption. 
Here, $n_{\rm e}$, $n_{\rm H}$, and $V$  are the number densities of electrons and protons, and the emitting volume, respectively.
The abundances relative to the solar values \citep{Anders89} of Ne, Mg, Si, S, Ar and Fe were allowed to vary freely, while the Ca and Ni abundances were linked to those of Ar and Fe, respectively.  
Abundances of the other elements were fixed to one solar.

The best-fit temperature of the CIE plasma model ($kT_{\rm e}$) was obtained to be $0.60^{+0.18}_{-0.02}$\,keV, but this model was statistically unacceptable ($\chi^2/\rm{dof}=2121/353$) with large residuals as shown in figure \ref{fig:spec}-a.  
The largest residual was found at 2.0\,keV, which corresponds to the Si Ly$\alpha$ emission. 
Therefore, it is suggested that the ionization state of this element is higher than that in the best-fit CIE state. 
In addition, the hump-like were found around 2.7\,keV and 3.5\,keV. 
These correspond to the K-shell binding energies of H-like Si (2.67\,keV) and H-like S (3.48\,keV).  
As is the case of IC\,443 \citep{Yamaguchi09}, these residuals are most likely to be RRCs in a recombining plasma (RP).

We thus introduced a single-component RP model, where $kT_{\rm e}$ and $kT_{\rm z}/kT_{\rm e}$ were allowed to be free parameters.
Then the hump-like structures in the residual and the excess at the Si Ly$\alpha$ were gone (see figure \ref{fig:spec}-b).
The best-fit $kT_{\rm e}$ and $kT_{\rm z}/kT_{\rm e}$ were $0.42\pm0.02$\,keV and $1.52\pm0.05$, respectively.  
However, the single RP model still failed to fit the overall spectrum with $\chi^2/\rm{dof}$ of 1175/352, with a large discrepancy between the data and model in the low-energy band ($<2.0$\,keV).  
Among the residual features, a ``dip'' at $\sim1.5$\,keV (shown with an arrow in figure \ref{fig:spec}-b) is most remarkable. 
This corresponds to the Mg-Ly$\alpha$ line, suggesting that the ionization temperature $kT_{\rm z}$ of Mg would be lower than those of Si and S. 
We supposed that the low-energy spectrum is dominated by a lower-temperature plasma attributable to the shocked ISM, and hence added a low-temperature CIE model. However this combined model 
(RP$+$CIE) gave no essential improvement in the low energy residuals.

Finally, we consider that each heavy element has an independent ionization temperature, as was first pointed out in IC\,443 \citep{Yamaguchi12}.
\citet{Sawada12} also found that $kT_{\rm z}$ for Ne and Mg in W28, another MM SNR, are closer to the CIE ($kT_{\rm z}/kT_{\rm e}$=1.2--1.7) than those for Si and S ($kT_{\rm z}/kT_{\rm e}$=2.4--2.5). 
They successfully demonstrated that this variation can be explained by the element-dependent recombination timescale (see also \cite{Smith10}).
We therefore tried a non-equilibrium ionization jump (NEIJ) model for W44.
This model describes a plasma state when the initial plasma is in CIE with the temperature of $kT_{\rm e1}$, then only the electron temperature dropped to $kT_{\rm e2}$ by a rapid electron cooling in the past. 
Assuming a constant $kT_{\rm e2}$,   the following evolution of the RP is  traced as a function of $n_{\rm e}t$, where $n_{\rm e}$ and $t$ are the number density of electrons and elapsed time, respectively\footnote{More detailed information about this model is found at http://www.sron.nl/files/HEA/SPEX/manuals/manual.pdf}

The result of the NEIJ model fit is shown in figure \ref{fig:spec}-c with the best-fit parameters of $kT_{\rm e1}=1.07^{+0.08}_{-0.06}$\,keV, $kT_{\rm e2}=0.48\pm0.02$\,keV, and log($n_{\rm e}t \ \rm{[s\,cm^{-3}]}\rm{)}=11.83\pm0.05$.
Although the residuals below $\sim2.0$\,keV were fairly improved ($\chi^2/\rm{dof}=928/351$), line-like excesses still remained at $\sim0.8$\,keV and $\sim1.2$\,keV (the arrows in figure \ref{fig:spec}-c).

As pointed out by a number of previous studies on SNRs, emissivity prediction of Fe L-shell lines are uncertain in currently available plasma code (e.g., \cite{Hughes98, Borkowski06}; \cite{Yamaguchi11}). 
\citet{Gu07} claimed that the line intensity ratio of Fe\emissiontype{XVII} corresponding to the 3s$\rightarrow$2p ($\sim730$\,eV) and 3d$\rightarrow$2p ($\sim820$\,eV) transitions may have a significant error.  
\citet{Brickhouse00} pointed out that several Fe L-shell lines (for example, $n\geq6$$\rightarrow$2 for Fe\emissiontype{XVII} and $n=6$, 7$\rightarrow$2 for Fe\emissiontype{XVIII}) are missing from the SPEX code.
We thus assumed  that the excesses at $\sim0.8$\,keV and $\sim1.2$\,keV are caused by such uncertainty, and added two Gaussians. 
Then the fit was improved with a $\chi^2/\rm{dof}$ value of 536/349 (figure \ref{fig:spec}-d). 
The best-fit parameters were $kT_{\rm e1}=1.25^{+0.11}_{-0.20}$\,keV, $kT_{\rm e2}=0.48\pm0.02$\,keV, log($n_{\rm e}t \ \rm{[s\,cm^{-3}]}\rm{)}=11.85^{+0.03}_{-0.06}$.
We note that the extra Gaussians give no significant change of the best-fit parameters of the simple NEIJ model (with no extra Gaussian). 
Hereafter, we used the NEIJ plus two Gaussians model, referring as the NEIJ model for simplicity.

\subsection{Fit with Thermal Plasma plus Power-Law}\label{sec:hard}
We tried the NEIJ model to the other two regions B and C. 
The model gave a nice fit for the region B spectrum, but failed to reproduce the region C spectrum with significant excess above $\sim4.0$\,keV.  
This excess is consistent with the hard band image (figure \ref{fig:image}-right;  5.0--7.0\,keV), where we found a previously unreported feature in the southwest part of the FoV.  
The hard component in the region C spectrum was found to be well explained with an additional power-law with a photon index $\Gamma$ of $2.12\pm0.11$ (figure \ref{fig:spec2}). 

We finally tried a combined model of  the NEIJ and power-law components for all the other regions (A and B), since we can see a hint of slight hard excess over the NEIJ model in the region A spectrum as well (see figure \ref{fig:spec}). 
The power-law index was fixed to the best-fit value for the region C ($\Gamma = 2.12$).  
The best-fit results are shown in figure \ref{fig:spec2} and table \ref{tab:spec}. 
We found that the power-low component is significant in the region A, while only an upper limit was obtained for the region B. 

\subsection{Spatial Distribution of the Recombining Plasma}\label{sec:ion}
\begin{figure*}[t]
  \begin{center}
  \FigureFile(170mm,170mm){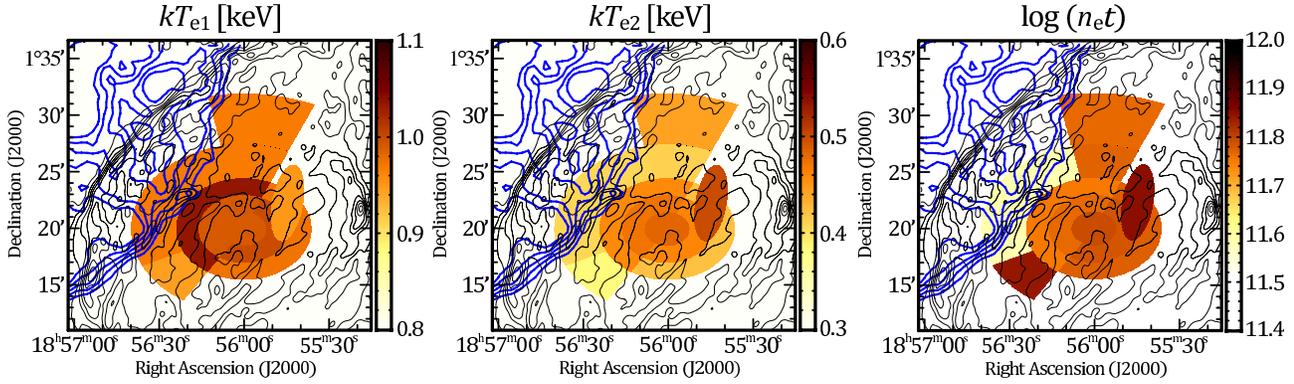}
  \end{center}
  \caption{Spatial distributions of the values of $kT_{\rm e1}$ (left), $kT_{\rm e2}$ (middle) and log($n_{\rm e}t$) (right). The black contours represent the 1442.5\,MHz radio continuum emission with VLA. The blue contours represent $^{12}$CO ($J=2$--1) in the velocity of 40-50\,km s$^{-1}$ obtained with NANTEN2 4\,m submillimeter telescope of Nagoya University.}\label{fig:map}
\end{figure*}

We further examined more detailed spatial structure of the recombining  plasma (RP) by dividing the entire FoV into 10 small regions and fitting each spectrum with the NEIJ plus power-law model.  
We fixed the photon index of the power-law to be 2.12 for all the regions. 
Statistically accepted results were obtained for all the spectra. 
Figure \ref{fig:map} shows spatial distributions of the best-fit parameters of $kT_{\rm e1}$ (left), $kT_{\rm e2}$ (middle) and log($n_{\rm e}t$) (right).
We found that the recombining state is achieved in the entire regions of the SNR. 

\section{Discussion}
\subsection{Recombining Plasma}
We found that the spectra from all the regions of W44 were nicely fitted with the model of the NEIJ and power-law components. 
In order to clarify the property of the plasma state, we estimate the average charge per element and compared to those expected in a CIE plasma with the same electron temperature. 
The results derived with the best-fit parameters for the region A spectrum are shown in table \ref{tab:average}.
We see that the averaged charge states of lighter elements, C, N, and O are consistent with those in CIE, while those of the heavier elements, such as  Si, S and Fe are higher than CIE, indicating that these elements are in the recombining state. 
In contrast, some previous X-ray studies claimed that the plasma in W44 is ionizing or nearly in CIE (\cite{Rho94}; \cite{Kawasaki05}).
One of the possible origins of this discrepancy is a difference in the energy resolution and/or photon statistics in the hard X-ray band.
%This is against the previous X-ray studies which claimed that the plasma in W44 is ionizing or nearly in CIE (\cite{Rho94}; \cite{Kawasaki05}).
%This discrepancy possibly originates from the difference in the energy resolution and/or photon statistics in the hard X-ray band.

Our result suggests that either  only the initial electron temperature $kT_{\rm e1}$ cooled down rapidly to $kT_{\rm e2}$ (here, electron cooling), or only the initial $T_{\rm z}$ (=$kT_{\rm e1}$) was raised higher than $kT_{\rm e2}$(here, selective ionization). 
These led the ions to be higher charge states (RP)  compared to those in CIE. 
In the next sections, we separately discuss these two cases.

\subsection{Electron Cooling}
One possible scenario for the electron cooling is thermal conduction, so we first evaluate this possibility. 
In this scenario, the energy transfer from the central region ($kT_e = 0.48 \pm 0.01$\,keV) to the cooler periphery ($kT_e = 0.40^{+0.02}_{-0.01}$\,keV) is considered. 
A scale length of the temperature gradient $l_{\rm T}$ = $({\rm grad} \, ln T)^{-1}$ is calculated to be $1.5\times10^{20}$\,cm, at the distance to the SNR of 3\,kpc. 
Therefore, the thermal conduction timescale $t_{\rm{cond}}$ is estimated to be 
\begin{eqnarray}
t_{\rm{cond}} \simeq 1\times10^7 \left(\frac{n_{\rm e}}{\rm1.0\,cm^{-3}}\right)\left(\frac{kT_{\rm e{\rm{(in)}}}}{0.48\, \rm{keV}}\right)^{-5/2} \nonumber \\
\times \left(\frac{l_{\rm T}}{\rm 1.5\times10^{20} \, cm}\right)^{2} \rm{yr}. \label{eq:t_cond}
\end{eqnarray}
(\cite{Spitzer62}; \cite{Kawasaki02}), 
where the electron number density $n_{\rm e}$ is calculated from the best-fit EM we obtained. 
The derived timescale is nearly 50 times longer than the age of W44 ($\sim 20,000$\,yr),  and hence the thermal conduction scenario is difficult to reproduce the observed electron temperature gradient.

The other possibility of the electron cooling is adiabatic expansion, which may occur in a rarefaction process; a rapid cooling of electrons occurs when a blast wave in a dense circumstellar medium (CSM) breaks into a lower-density ISM (\cite{Itoh89}; \cite{Shimizu12}; \cite{Moriya12}). 
\citet{Sawada12} argued the recombination time $t_{\rm rec}$ is an essential parameter to evaluate the possibility of this scenario.
We hence estimate the $t_{\rm rec}$ using the best-fit $n_{\rm e}t$ values: the lowest  at the east region of $3.8\pm0.1\times10^{11}$\,s\,cm$^{-3}$, and the average in the center region of $\sim7.1\pm0.1\times10^{11}$\,s\,cm$^{-3}$ (figure \ref{fig:map}-c).
Since the electron number densities $n_{\rm e}$ at the center and east region are, respectively $\sim1.0$\,cm$^{-3}$ and $\sim0.8$\,cm$^{-3}$, the recombining time is derived to be $\sim$15,000--20,000\,yr.
\citet{Wolszczan91} estimated the age of the pulsar PSR\,B1853$+$01, physically associated with W44, to be 20,000\,yr, which is consistent with our estimated recombination time.  
Therefore, the adiabatic cooling scenario is preferable in our case, because the shock-heating and the rarefaction should occur at the initial phase of the SNR evolution.
\citet{Broersen11} derived conditions for general SNRs to keep RP in their evolutions.
Comparing the $t_{\rm rec}$ with an adiabatic cooling time $t_{\rm ad}$ in different $n_{\rm e}$, they concluded that a low $n_{\rm e}$ significantly decreases the recombination rate.
They claimed SNR\,0506$-$68 ($\sim4,000$\,yr) to be still RP in a dense ISM ($n_{\rm e}\sim10$\,cm$^{-3}$).
Applying their argument to W44's case, we confirmed that the $t_{\rm rec}$ for W44 is about one order of magnitude above that for SNR\,0506$-$68 (see also \cite{Vink12}).
This estimation also supports the adiabatic cooling scenario.

\begin{table}[t]
  \caption{Average Charge Per Element}\label{tab:average}
  \begin{center}
    \begin{tabular}{ccc}
       \hline 
      \hline
Element & \multicolumn{2}{c}{Average Charge} \\
 & NEIJ (Region A) & CIE at 0.45\,keV \\
      \hline
C      	&    		5.99 	&      	5.99 \\
N      	&          	6.98 	&      	6.98 \\
O       	&          	7.92 	&      	7.92 \\
Ne     	&           	9.51 	&        	9.34\\
%Na     	&           	10.21 	&    		9.764\\
Mg    	&          	10.87 	&        	10.32\\
%Al       	&          	11.55 	&       	11.11\\
Si       	&         	12.26 	&        	12.01\\
S       	&         	13.95 	&       	13.91\\
%Ar      	&           	15.66 	&      	15.68\\
%Ca     	&           	16.89 	&      	16.76\\
Fe     	&           	16.45 	&      	16.37\\
%Ni      	&            	18.72 	&     		18.61\\
      \hline
    \end{tabular}
 \end{center}
\end{table}

\subsection{Selective Ionization}
Possible mechanisms to rise up $kT_{\rm z}$ (selective ionization)  would be either ionization by supra-thermal electrons (e.g., \cite{Tanaka86}; \cite{Masai02}), or photo-ionization (e.g., \cite{Piro00}).  
In W44, the 2--10\,GeV and  0.4--3\,GeV emissions were discovered with the Fermi LAT \citep{Abdo10} and AGILE \citep{Giuliani11}, respectively.  
The idea of the former scenario is that supra-thermal particles are injected into a Fermi acceleration process, and are accelerated up to very high energy. 
The supra-thermal particles also ionize atoms in a thermal plasma to higher ionized states, and produce the RP.  
In fact, large solar flares often show a signature of RP (in Fe atoms) and hard X-ray emissions \citep{Tanaka86}. 
Thus, the scenario of supra-thermal particle ionization may predict positive correlation between the ionization temperature ($kT_{\rm z}$) and the flux of TeV/GeV $\gamma$-rays/hard X-rays.  
However, our observation shows no systematic spatial variation of $kT_{\rm z}$, and hence possibility of selective ionization by supra-thermal particles is unlikely.

Another possibility, photo-ionization, requires a strong X-ray source in or near W44, but none of such sources is found at present. 
Therefore, such ionizing source should, if any, be bright in past.
A $\gamma$-ray burst is one of the candidates.
If W44 was a remnant of hyper-nova explosion associated with a $\gamma$-ray burst, the X-ray afterglow would be strong enough to ionize a substantial amount of atoms (high $kT_{\rm z}=kT_{\rm e1}$), whereas the electron temperature was relatively low (low $kT_{\rm e2}$).  
A weak point of this scenario is, however, that no particular signature for a hyper-novae has so far been observed in W44.

\subsection{Origin of Hard X-ray Emission}\label{sec:hardxray}
We detected strong hard X-rays from the west region of W44 (region C), and possibly from the entire SNR regions.
A conceivable origin is a pulsar wind nebula (PWN) associated with PSR\,B1853$+$01 (\cite{Wolszczan91}; \cite{Petre02}) located just outside the FoV (the white cross in figure \ref{fig:image}).
The Chandra ACIS found that the X-ray emitting region of the PWN is within $\sim\timeform{2'}$ from PSR\,B1853$+$01, while we found that the peak of the hard X-rays is far from the PWN ($\sim\timeform{12'}$).
The total flux of the PWN plus PSR\,B1853$+$01 is $\sim1.7\times10^{-13}$\,ergs\,cm$^{-2}$\,s$^{-1}$ in the 2--8\,keV band \citep{Petre02}, far less than the hard X-ray flux in the region C ($\sim1.37\times10^{-12}$\,ergs\,cm$^{-2}$\,s$^{-1}$).
We also simulated possible contamination of the strayed flux from the PWN and PSR\,B1853$+$01 to the region A by using \textit{xissim} \citep{Ishisaki07}, and confirmed it is negligible. 
We thus conclude that the hard X-ray emission we discovered is independent from the PWN or PSR\,B1853$+$01.

\begin{figure}[!t]
  \begin{center}
    \FigureFile(70mm,70mm){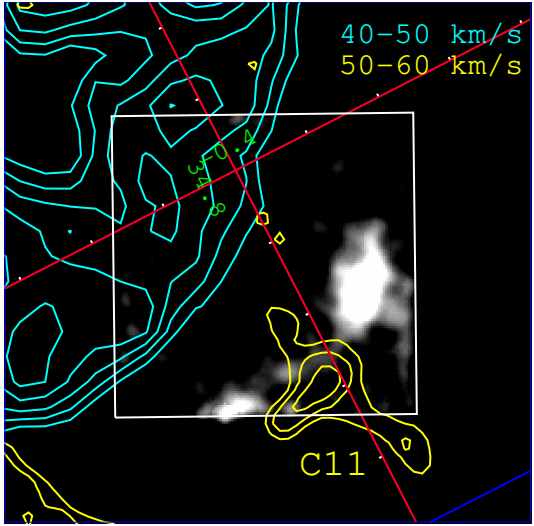}
  \end{center}
  \caption{$^{12}$CO ($J=2$--1) contours of NANTEN2 overlaid with the hard X-ray image in the 5.0--7.0\,keV band. The blue and yellow contours are the  $^{12}$CO fluxes in the velocity ranges 40--50\,km\,s$^{-1}$ and 50--60\,km\,s$^{-1}$, respectively. The white square is the XIS FoV.}\label{fig:co}
\end{figure}

The hard X-rays was found to have an arc-like structure (figure \ref{fig:image}-right) and are correlated with a filament of the radio continuum (1442.5\,MHz with VLA).
For more detailed study, we compared the spatial distributions of the hard X-rays with ambient molecular clouds, $^{12}$CO ($J=2$--1) emission obtained by NANTEN2.
The result is shown in figure \ref{fig:co}. 
We see a clear anti-correlation between the hard X-rays and the CO cloud in the velocity range 50--60\,km\,s$^{-1}$ (the yellow contours).
This cloud is identical to what was observed by Nobeyama 45-m radio telescope, C11 named by \citet{Seta04}, and was suggested to be interacting with the SNRs blast wave  (see figure 11 in their paper).

The arc-like structure of the hard X-rays suggests that this emission comes from the shock.
A typical shock velocity of W44 is $\sim650\,(kT_{\rm e}/0.5\,\rm{keV})^{0.5}$\,km\,s$^{-1}$.
As far as we assume a simple direct diffusive shock acceleration, such slow shock is insufficient to explain the flux of the observed hard X-rays.
We presume that the flux enhancement possibly arises in association with the shock-cloud interaction.
We focus on pre-existing cosmic ray (CR) electrons compressed and reaccelerated between the shock and dense cloud.
In the radio band, the resultant synchrotron enhancement is known as the van der Laan mechanism (\cite{Laan62a}; 1962b) which accounts for the radiative radio shell \citep{Blandford82}. 
Based on a multi-wavelength observation of W44, \citet{Cox99} pointed out that the radio synchrotron emission is possibly enhanced by the pre-existing CR electrons which are reaccelerated by the van der Laan mechanism.
\citet{Uchiyama10} applied a similar mechanism to some middle-aged SNRs including W44 to interpret their GeV $\gamma$-ray emission. 
They concluded that it is sufficient for the reaccelerating of the pre-existing CR through the hadronic process to explain the observed GeV $\gamma$-ray intensity.
We simply expanded their argument to explain the observed hard X-rays.
By our rough simulation, if the hard X-ray is derived from such mechanism, the electrons shall be accelerated to at least $\sim$100\,TeV.
In this case, the acceleration time $t_{\rm acc}$ is calculated to be
\begin{eqnarray}
t_{\rm{acc}} \simeq 1\times10^6 \left(\frac{cp_{\rm max}}{\rm100\,TeV}\right)\left(\frac{\eta}{10}\right)\left(\frac{v_{\rm s}}{\rm 100\,km\,s^{-1}}\right)^{-2} \nonumber \\
\times \left(\frac{B_{\rm 0}}{\rm 25\, \mu \rm{G}}\right)^{-1} \rm{yr}, \label{eq:t_acc}
\end{eqnarray}
where $cp_{\rm max}$ is the maximum attainable energy, $\eta$ is the gyrofactor \citep{Uchiyama07}, $v_{\rm s}$ and $B_{\rm 0}$ are the shock velocity decelerated by cloud and the pre-shock magnetic field \citep{Reach05}, respectively.
These parameters were set as \citet{Uchiyama10} had assumed for W44 in their calculation.
As a result, the $t_{\rm{acc}}$ becomes significantly longer than the age.

A local anti-correlation between CO clouds and non-thermal X-ray emissions is also found in another SNR, RX\,J1713.7$-$3946 \citep{Sano10}.
Performing magnetohydrodynamic (MHD) simulations of a SNR shock passing through clumpy interstellar clouds, \citet{Inoue12} successfully interpreted this result; the synchrotron X-ray enhancement occurs in spatially small regions at the vicinity of the cloud due to an amplified magnetic field caused by turbulent dynamo action (see also, \cite{Giacalone07}).
One major difference between RX\,J1713.7$-$3946 and W44 is a shock velocity.
\citet{Inoue10} showed that a slow shock wave ($\sim500$\,km\,s$^{-1}$) attains a maximum magnetic field strength of  $\sim400$\,$\mu$G.
Given a typical shock velocity of W44, $\sim650$\,km\,s$^{-1}$, however, the resultant $t_{\rm acc}$ becomes substantially longer than the synchrotron cooling time $t_{\rm syn}$; $t_{\rm syn}/t_{\rm acc}<0.07$.
We concluded that this scenario is also difficult to explain the observed X-ray flux.

In either case, the acceleration time $t_{\rm acc}$ is too long to obtain TeV electrons.
Alternatively, if TeV protons are still sufficiently remaining in W44, the hadronic process provides secondary electrons which may peak at TeV energies.
As already claimed by \citet{Abdo10} and \citet{Giuliani11}, the GeV $\gamma$-ray in W44 is possibly due to the hadronic process.
We found no spatial correlation between the hard X-ray emitting region and the global GeV $\gamma$-ray distribution (\cite{Abdo10}; \cite{Giuliani11}). 
The reason would be that the synchrotron X-ray is enhanced only around a shocked clumpy cloud where the magnetic field is amplified by the mechanism of \citet{Inoue10}. 
The remaining problem is a CR escape which determines a peak energy of the existing protons, hence the secondary electrons in W44.
While the maximum possible energy differs by more than one order of magnitude among theories (e.g., \cite{Ptsukin03}), it is generally difficult for a middle-aged SNR to have 100-TeV order of protons.
The origin of the hard X-rays remains an open question.

\section{Summary}
\begin{enumerate}
\item The thermal X-ray emission in W44  is  centrally-filled,  similar to other MM SNRs, while the arc-like hard X-ray emission has been found for the first time.
\item The thermal X-ray spectra are well represented with a recombining plasma (NEIJ model).
\item The electron temperature of the thermal plasma would be cooled down rapidly from $\sim1$\,keV to 0.4--0.5\,keV at $\sim$20,000 years ago.
\item The most plausible origin of the recombining plasma is rarefaction in the early phase of the SNR evolution.
\item The surface brightness of the hard X-ray and $^{12}$CO ($J=2$--1) emissions are spatially anti-correlated.
\item Some particle acceleration mechanisms which we discussed give no plausible explanation of the anti-correlation. The origin of the hard X-ray emission is an open question.
%Plausible explanation of the anticorrelation is that the blast wave is interacting with a small molecular cloud so that the magnetic field at the periphery are amplified to enhance the hard X-ray flux.
\end{enumerate}

\section*{Acknowledgments}
H.U., M.S. and T.O. are supported by Japan Society for the Promotion of Science (JSPS) Research Fellowship for Young Scientists. 
H.Y. is supported by JSPS Research Fellowship for Research Abroad.
%The work is partially supported by the Ministry of Education, Culture, Sports, Science and Technology (Grant-in-Aid No.23000004).

\end{document}